\documentclass{PoS}

\graphicspath{ {figs_vasp_im/} }

\title{Massive quark form factors}

\ShortTitle{Massive quark form factors}

\author{Roman N. Lee\\
        Budker Institute of Nuclear Physics\\
        630090 Novosibirsk, Russia\\
        E-mail: \email{r.n.lee@inp.nsk.su}}

\author{Alexander V. Smirnov\\
        Research Computing Center, Moscow State University\\
        119991, Moscow, Russia\\
        E-mail: \email{asmirnov80@gmail.com}}

\author{Vladimir A. Smirnov\\
        Skobeltsyn Institute of Nuclear Physics of Moscow State University\\
        119991, Moscow, Russia\\
        E-mail: \email{smirnov@theory.sinp.msu.ru}}

\author{\speaker{Matthias Steinhauser}\\
        Institut f{\"u}r Theoretische Teilchenphysik,\\ Karlsruhe
        Institute of Technology, 76128 Karlsruhe, Germany\\
        E-mail: \email{matthias.steinhauser@kit.edu}}

\abstract{In this contribution we provide further details to the recent calculation
  of QCD corrections to massive three-loop form factors presented in Ref.~\cite{Lee:2018rgs}.}

\FullConference{Loops and Legs in Quantum Field Theory (LL2018)\\
		29 April 2018 - 04 May 2018\\
		St. Goar, Germany}

\begin{document}

\section{Introduction}

Form factors play an important role in any quantum field theory. They are
building blocks for various physical quatities and furthermore provide a
playground for studying infrared properties.
In this contribution form factors involving massive quark are discussed.
The most advanced results have been obtained in Refs.~\cite{Lee:2018rgs}
and~\cite{Ablinger:2018yae}, where three-loop corrections have been considered.
More precisely, all light-fermion contributions have been computed and the
non-fermionic part has been considered in the large-$N_c$ limit.
In the latter case only planar Feynman diagrams contribute whereas in the
former case also non-planar integral families have to be taken into account.
For references to previous work, in particular to two-loop calculations, we
refer to~\cite{Lee:2018rgs}.

In this contribution we provide more details to the calculation performed in
Ref.~\cite{Lee:2018rgs}. Furthermore, numerical results are presented for the
imaginary part of the form factors.

\section{Quark form factors}

We consider QCD correction to the interaction of massive quarks with
a vector, axial-vector, scalar or pseudo-scalar current defined as
\begin{eqnarray}
  j_\mu^v &=& \bar{\psi}\gamma_\mu\psi\,,\nonumber\\
  j_\mu^a &=& \bar{\psi}\gamma_\mu\gamma_5\psi\,,\nonumber\\
  j^s &=& m \,\bar{\psi}\psi\,,\nonumber\\
  j^p &=& i m \,\bar{\psi}\gamma_5\psi\,.
  \label{eq::currents}
\end{eqnarray}
The tensor decomposition of the corresponding vertex functions leads to six
scalar functions which we define as
\begin{eqnarray}
  \Gamma_\mu^v(q_1,q_2) &=& 
  F_1^v(q^2)\gamma_\mu - \frac{i}{2m}F_2^v(q^2) \sigma_{\mu\nu}q^\nu
  \,, \nonumber\\
  \Gamma_\mu^a(q_1,q_2) &=& 
  F_1^a(q^2)\gamma_\mu\gamma_5 {- \frac{1}{2m}F_2^a(q^2) q_\mu }\gamma_5
  \,, \nonumber\\
  \Gamma^s(q_1,q_2) &=& {m} F^s(q^2)
  \,, \nonumber\\
  \Gamma^p(q_1,q_2) &=& {i m} F^p(q^2) {\gamma_5}
  \,.
  \label{eq::Gamma}
\end{eqnarray}
$F_1^v, \ldots, F^p$ depend on the ratio of the virtuality $q^2$, where $q$ is
the outgoing momentum of the external current, and the square of the heavy
quark mass $m$. For the practical calculation it is convenient to introduce
the variable $x$ defined through
\begin{eqnarray}
  \frac{q^2}{m^2} &=& - \frac{(1-x)^2}{x}
  \,,
  \label{eq::trans_x_q}
\end{eqnarray}
which maps the complex $q^2/m^2$ plane into the unit circle. In particular, we
have that the special points $q^2=0$, $q^2=4m^2$ and $q^2\to\-\infty$ are
mapped to $x=1,-1$ and $0$, respectively. 

\begin{figure}[t] 
  \begin{center}
    \begin{tabular}{cccc}
      \\[.3em]
      \includegraphics[width=0.2\textwidth]{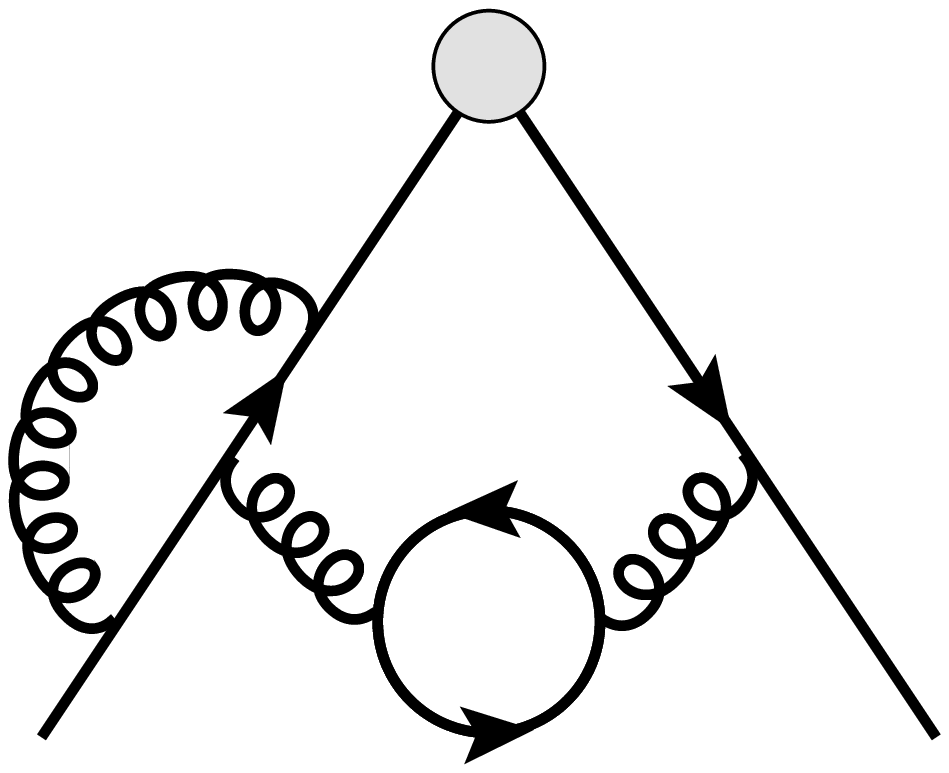} &
      \includegraphics[width=0.2\textwidth]{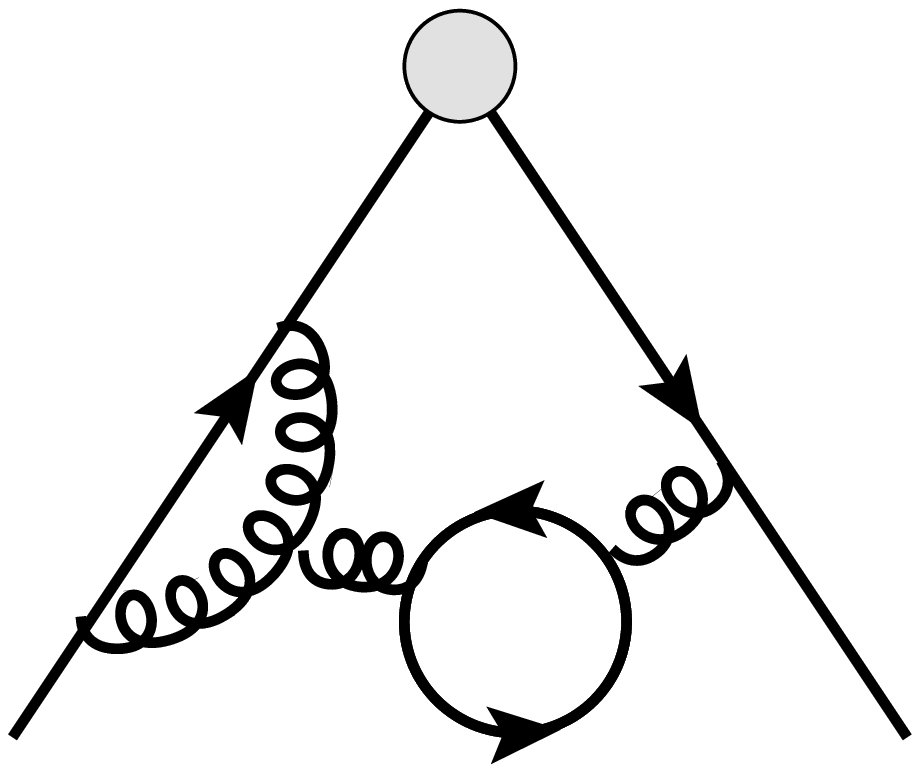} &
      \includegraphics[width=0.2\textwidth]{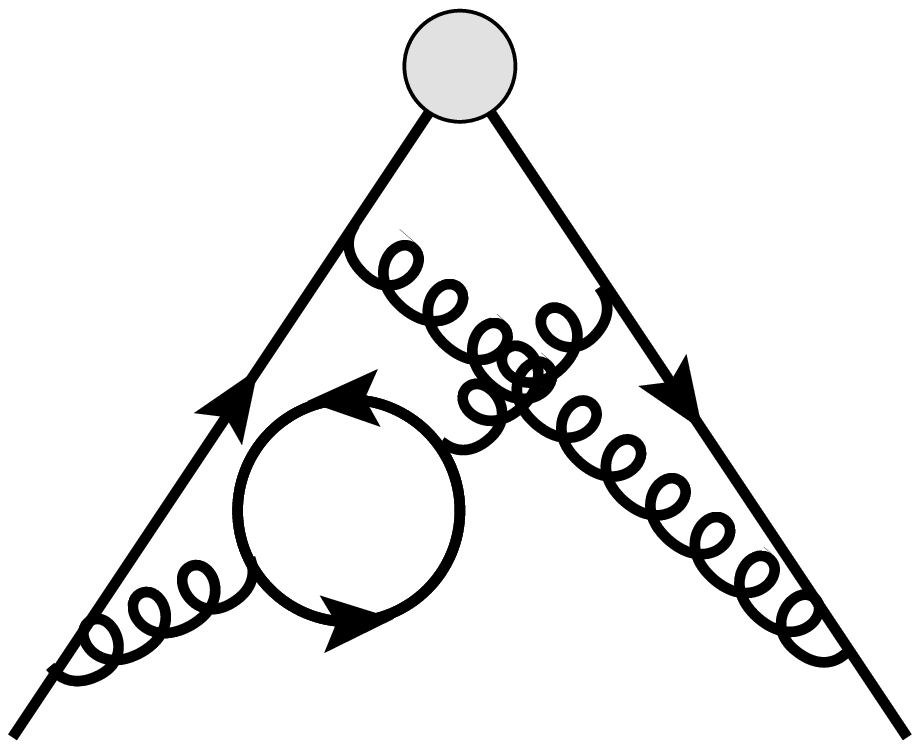} &
      \includegraphics[width=0.2\textwidth]{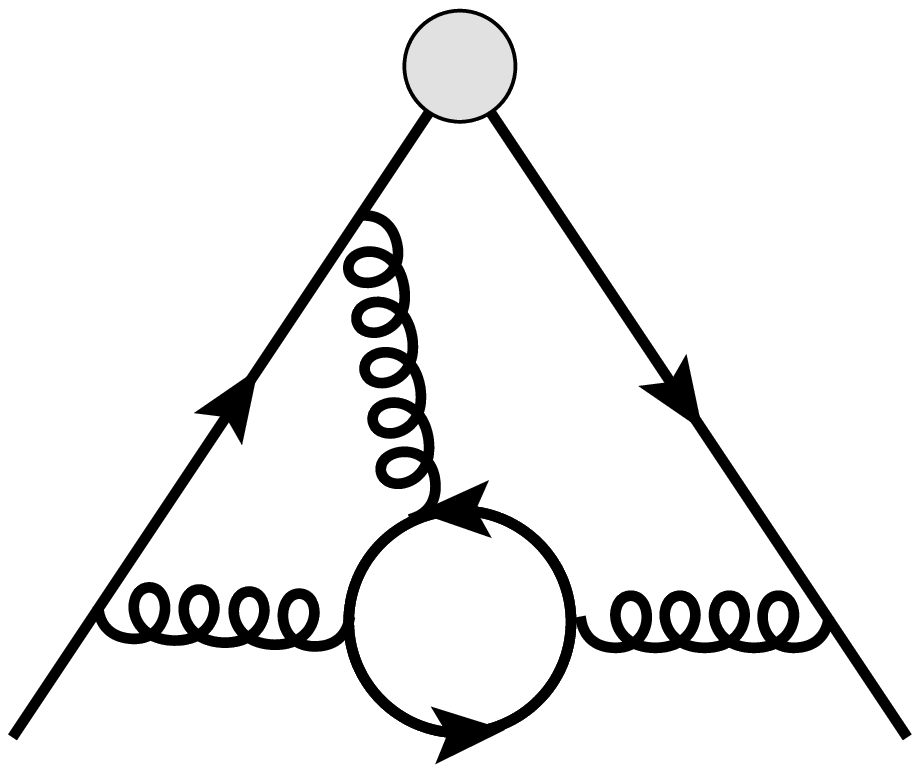}
      \\
      \includegraphics[width=0.2\textwidth]{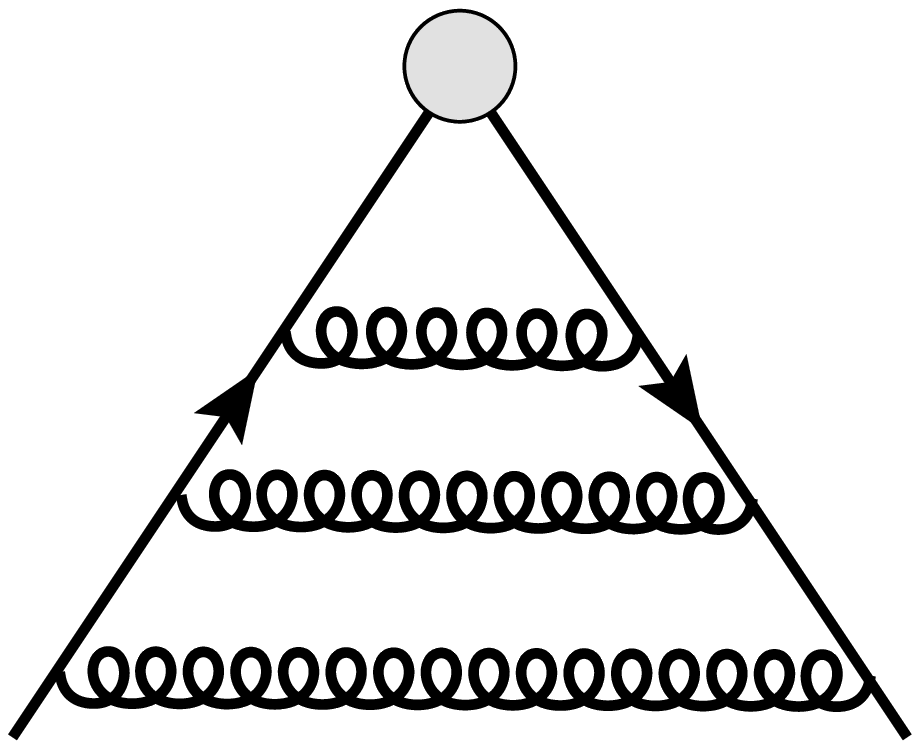} &
      \includegraphics[width=0.2\textwidth]{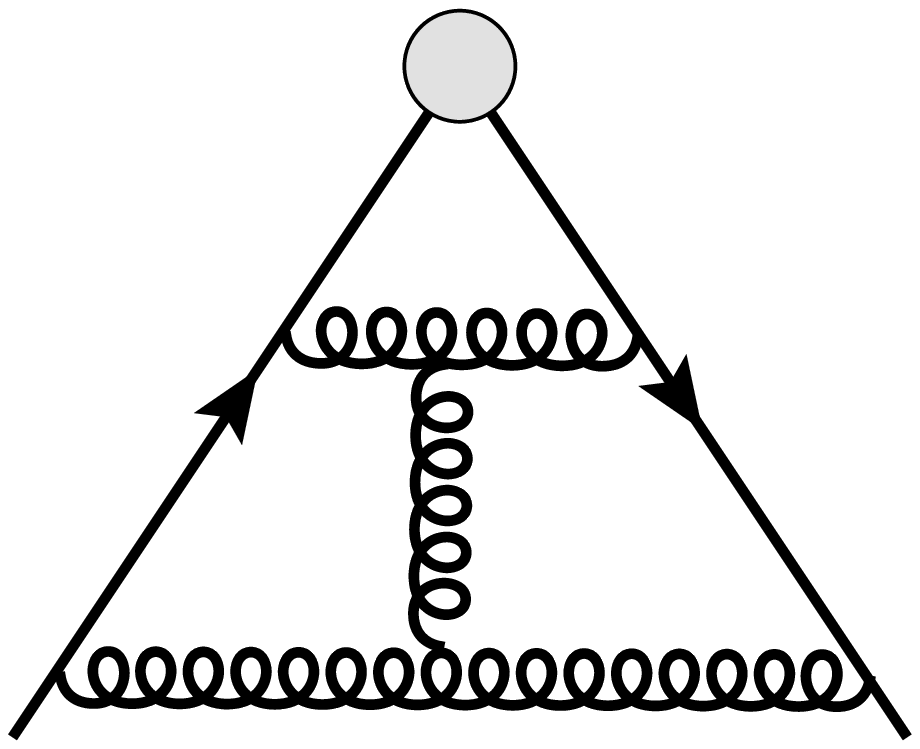} &
      \includegraphics[width=0.2\textwidth]{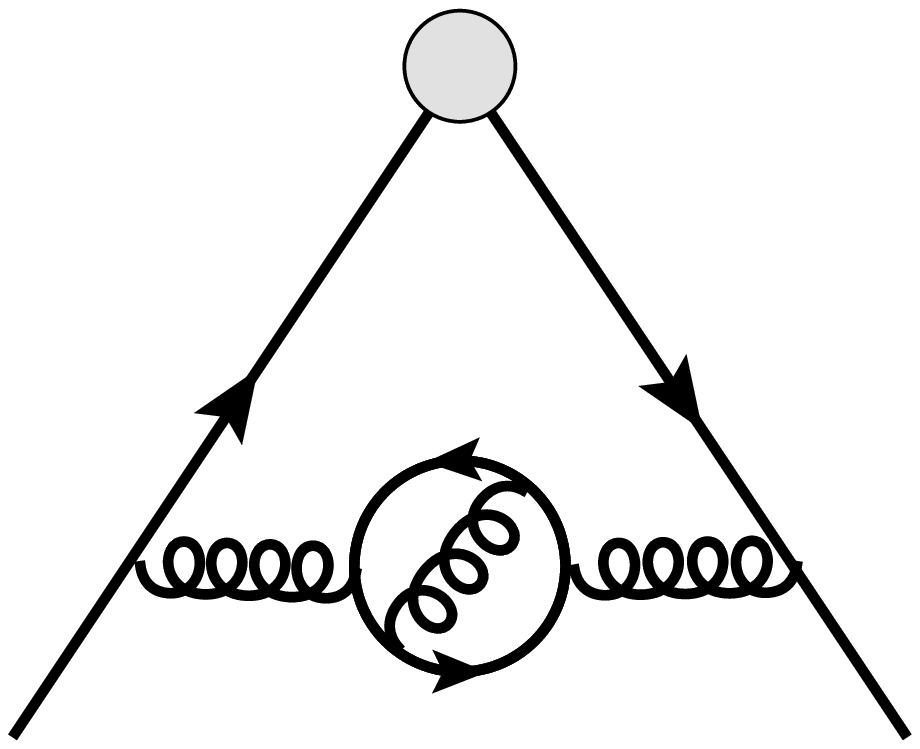} &
      \includegraphics[width=0.2\textwidth]{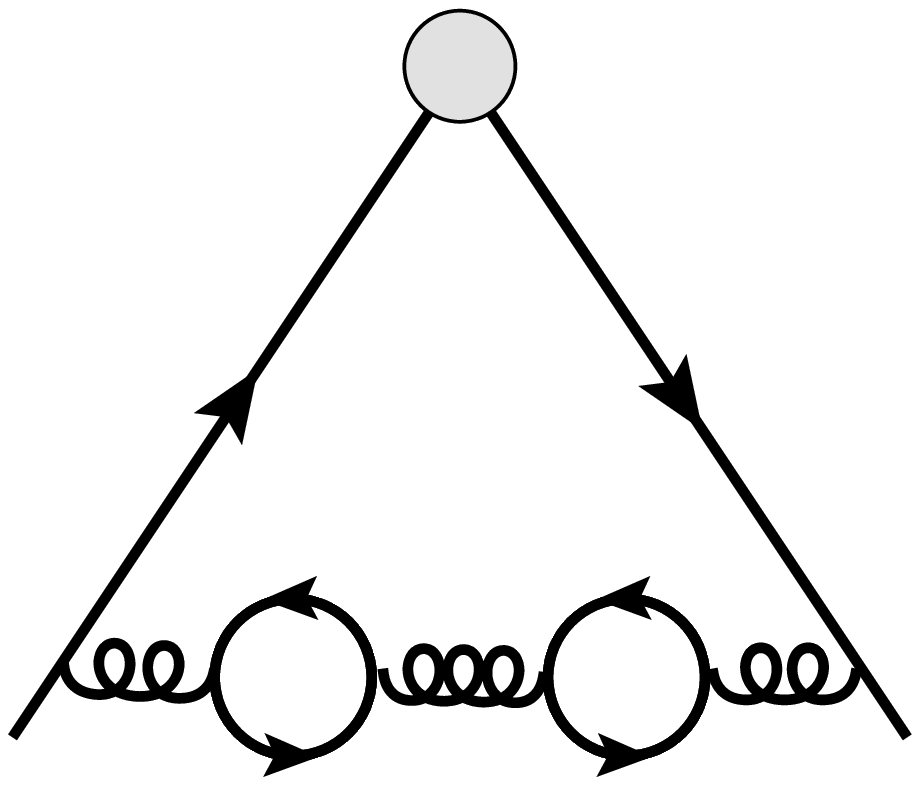}
    \end{tabular}
    \caption{\label{fig::diags}Sample diagrams contributing to the form
      factors.  Solid and curly lines represent quarks and gluons,
      respectively. The grey blob refers to one of the external currents given
      in Eq.~(\ref{eq::currents}).}
  \end{center}
\end{figure}

The workflow of our calculation is as follows: We generate the amplitudes
using {\tt qgraf}~\cite{Nogueira:1991ex}. Althogether 337~diagrams are
generated, which, however, also includes non-planar integrals. We use {\tt
  color} to compute the colour factors which allows us to select the
fermionic and large-$N_c$ contributions. Representative three-loop diagrams
can be found in Fig.~\ref{fig::diags}. We use {\tt
  q2e}~\cite{Harlander:1997zb,Seidensticker:1999bb} to transform the {\tt
  qgraf} output to {\tt FORM}~\cite{Ruijl:2017dtg} notation and {\tt
  exp}~\cite{Harlander:1997zb,Seidensticker:1999bb} together with the
underlying symmetries of the vertex diagrams to map all contributing integrals
to eight planar and three non-planar integral families (see Fig.~1 of
Ref.~\cite{Henn:2016kjz} and Fig.~5 of Ref~\cite{Lee:2018rgs} for graphical
representations).

In a next step we compute the amplitudes of each Feynman diagram using {\tt
  FORM}~\cite{Ruijl:2017dtg}. We apply projectors to obtain the scalar
function $F_1^v, \ldots, F^p$, take traces and map each integral to a scalar
function as defined by the corresponding integral family.  For most diagrams
this step takes only a few minutes for a general QCD gauge parameter. Afterwards
we extract the integral list which serves as input for {\tt
  FIRE}~\cite{Smirnov:2014hma}. {\tt FIRE} is used in combination with {\tt
  LiteRed}~\cite{Lee:2012cn,Lee:2013mka} for the reduction to master
integrals. For the large-$N_c$ contribution we have ${\cal O}(10^5)$ and for
the $n_l$ term about $30\,000$ integrals.  For the most complicated integral
family the reduction to master integrals takes of the order of a week on a
12-core node with main memory of order 100 gigabyte.

Analytic results for the master integrals are available from
Refs.~\cite{Henn:2016kjz} (89 planar integrals) and~\cite{Lee:2018nxa}
(additional 15 integrals, two of them are non-planar), respectively. They are
used to express each of the six form factors $F_1^v, \ldots, F^p$ in terms of
Goncharov polylogarithms (GPLs)~\cite{Goncharov:1998kja}. Computer-readable
expressions can be found in the ancillary file to Ref.~\cite{Lee:2018rgs}. They are
quite lengthy, however, a numerical evaluation is possible with the help of
{\tt ginac}~\cite{Bauer:2000cp,Vollinga:2004sn}.

There are several checks on the correctness of our result: First of all, we
have verified that in the renormalized form factors the QCD gauge parameter
drops out. We also obtain the expected limiting behaviours for $x\to0,1,-1$
(see Ref.~\cite{Lee:2018rgs} for an extensive discussion).  Furthermore, we
can check the pole part against a dedicated calulation of the cusp anomalous
dimension $\Gamma_{\rm cusp}$ which has been performed in
Ref.~\cite{Grozin:2015kna}.  Note that we obtain the same result for
$\Gamma_{\rm cusp}$ for all four currents of Eq.~(\ref{eq::currents}) which is
expected due to the universality of the infra-red behaviour.  Finally, we
obtain complete agreement for the renormalized three-loop form factors $F_1^v,
\ldots, F^p$ with an independent calculation performed in
Ref.~\cite{Ablinger:2018yae}.  Let us stress that in~\cite{Ablinger:2018yae}
different software has been used to perform the reduction to master integrals,
the integral families are defined differently and a different method has been
used to compute the master integrals.


\section{Numerical results}

\begin{figure}[t] 
  \begin{center}
    \begin{tabular}{ccc}
      \includegraphics[width=0.3\textwidth]{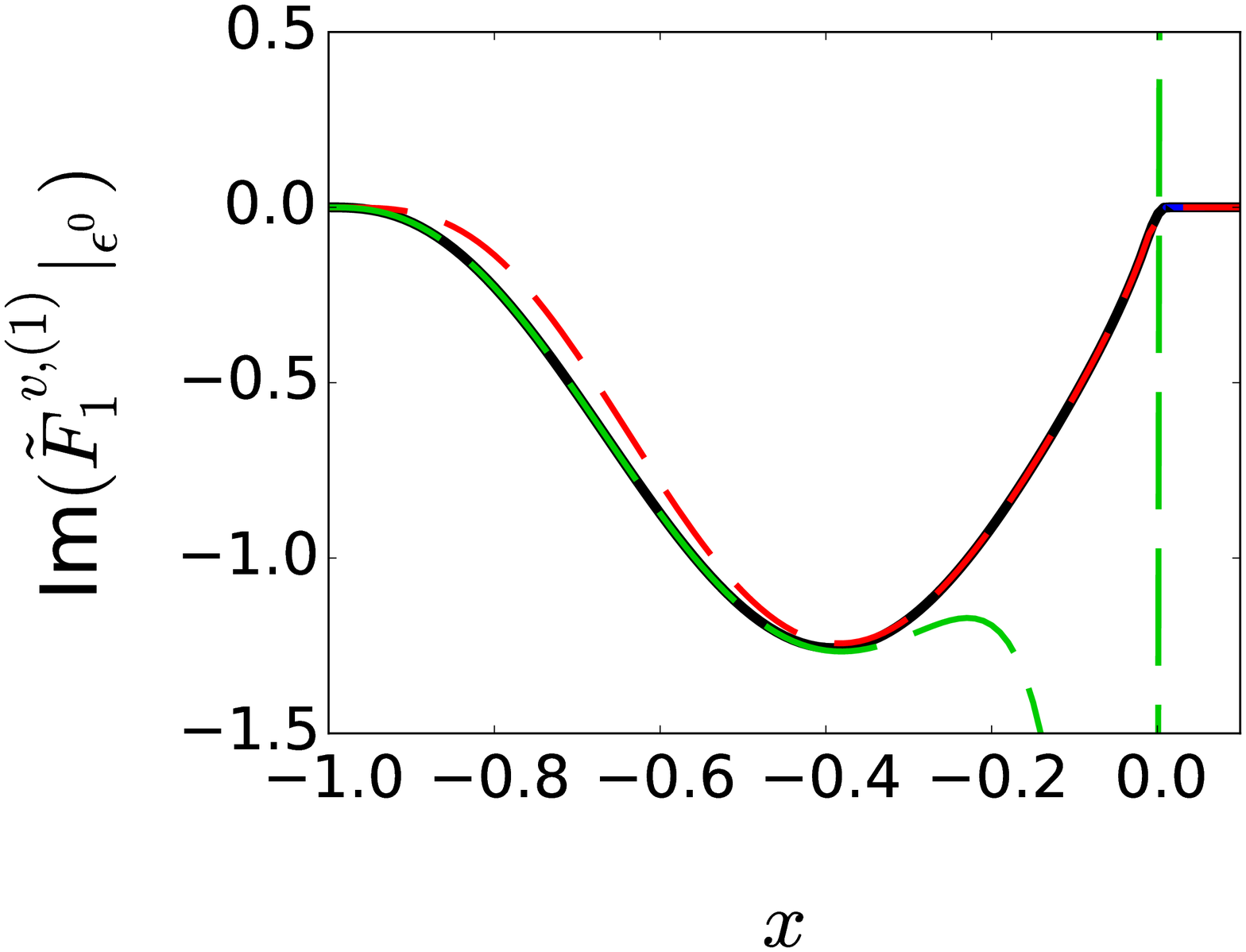} &
      \includegraphics[width=0.3\textwidth]{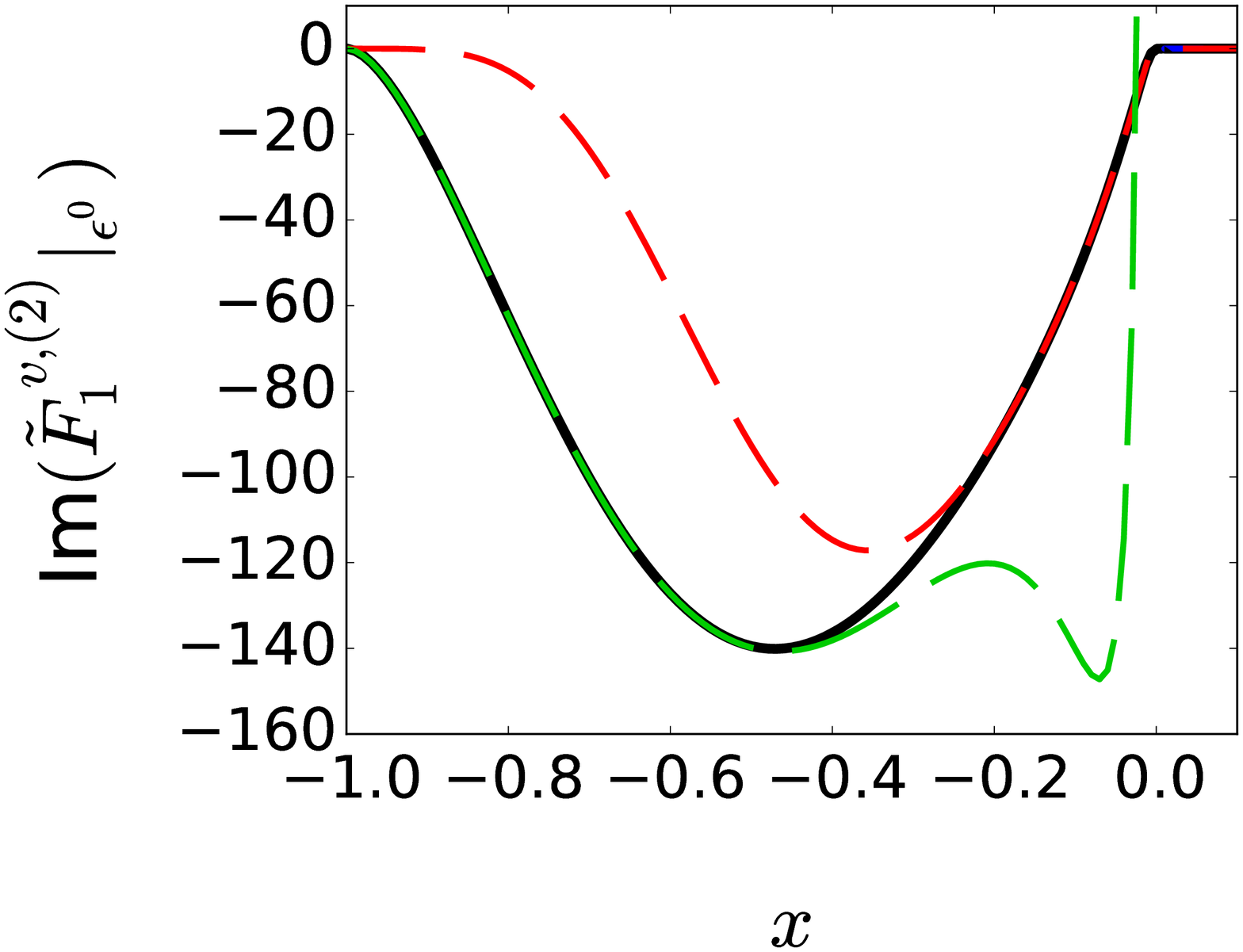} &
      \includegraphics[width=0.3\textwidth]{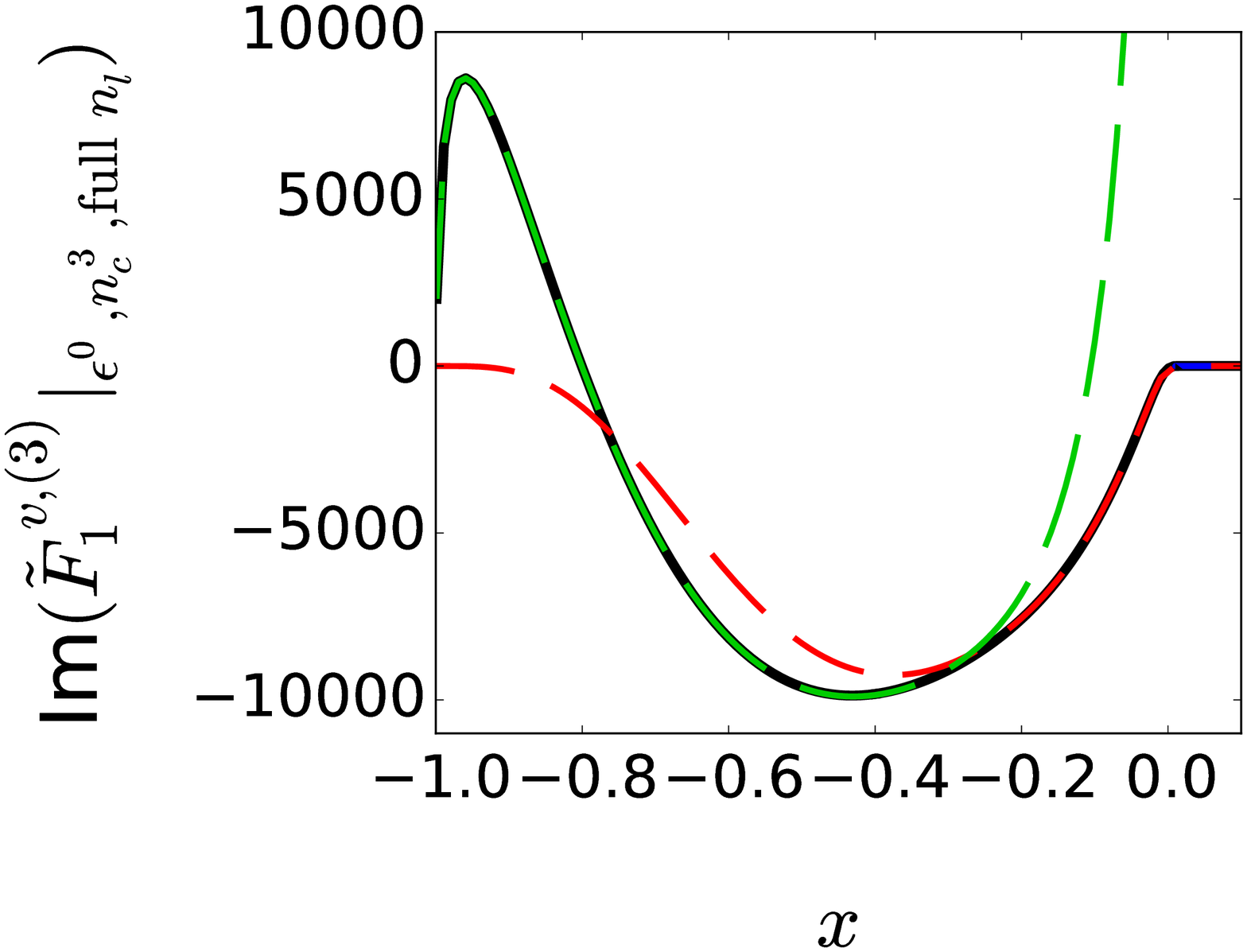} 
      \\
      \includegraphics[width=0.3\textwidth]{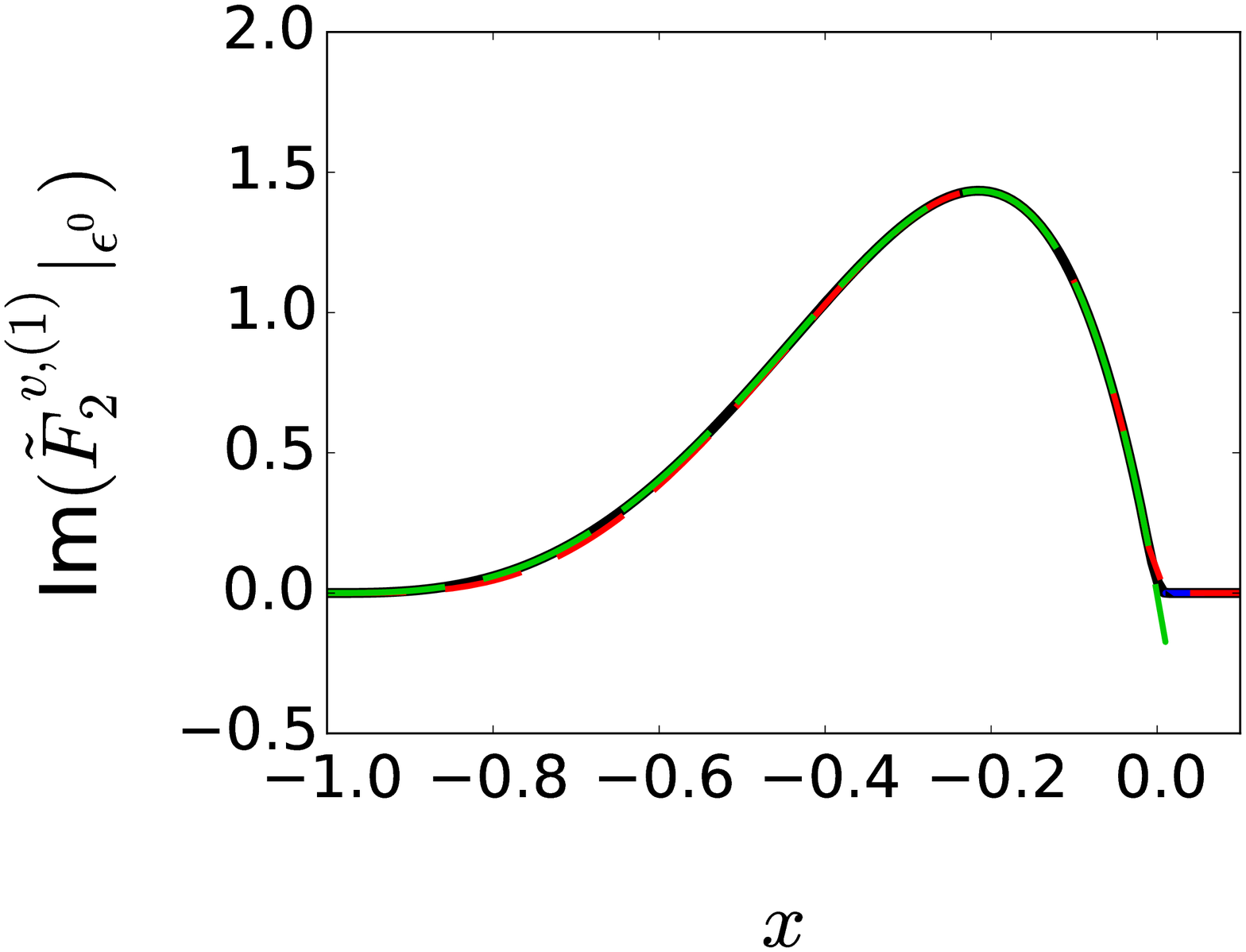} &
      \includegraphics[width=0.3\textwidth]{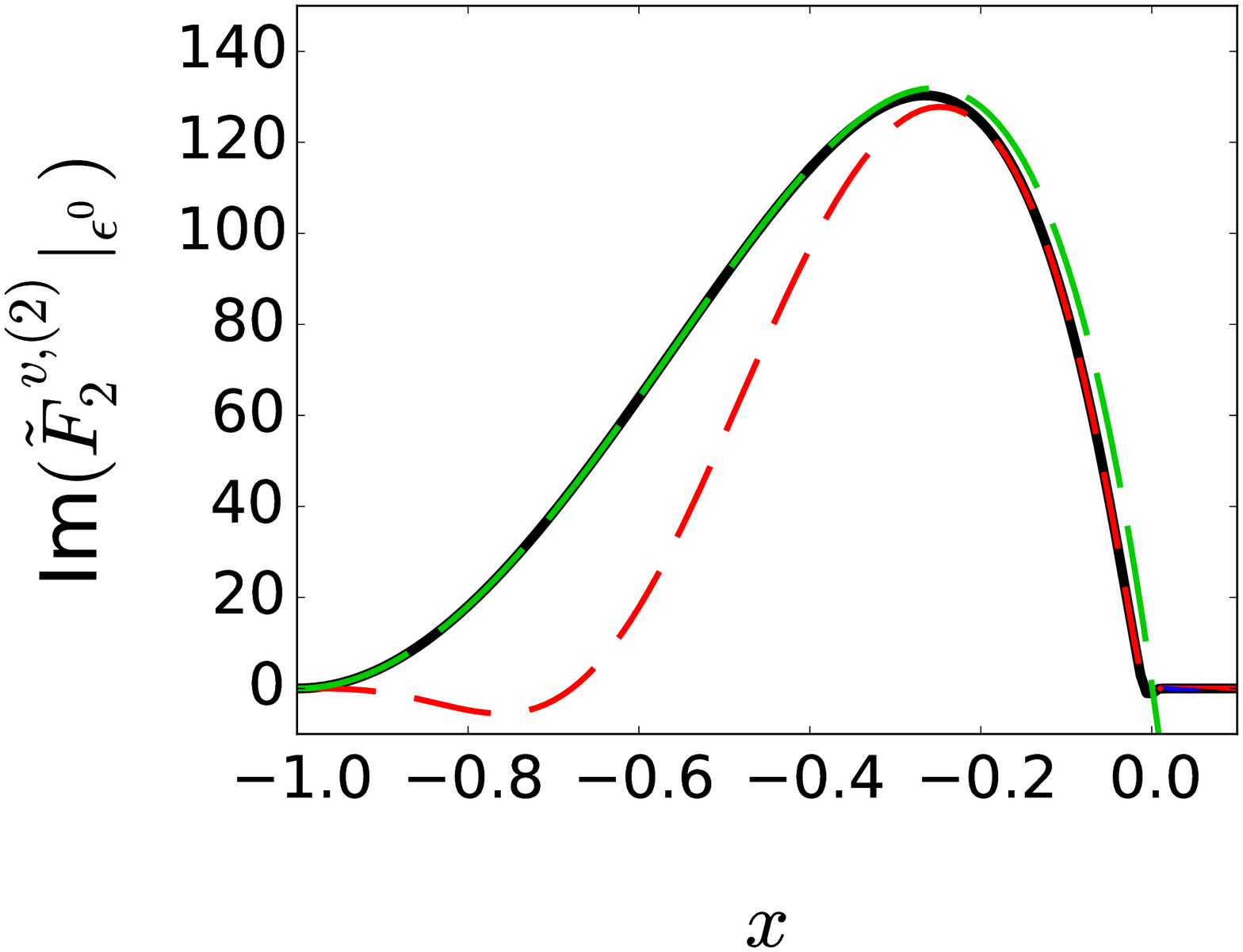} &
      \includegraphics[width=0.3\textwidth]{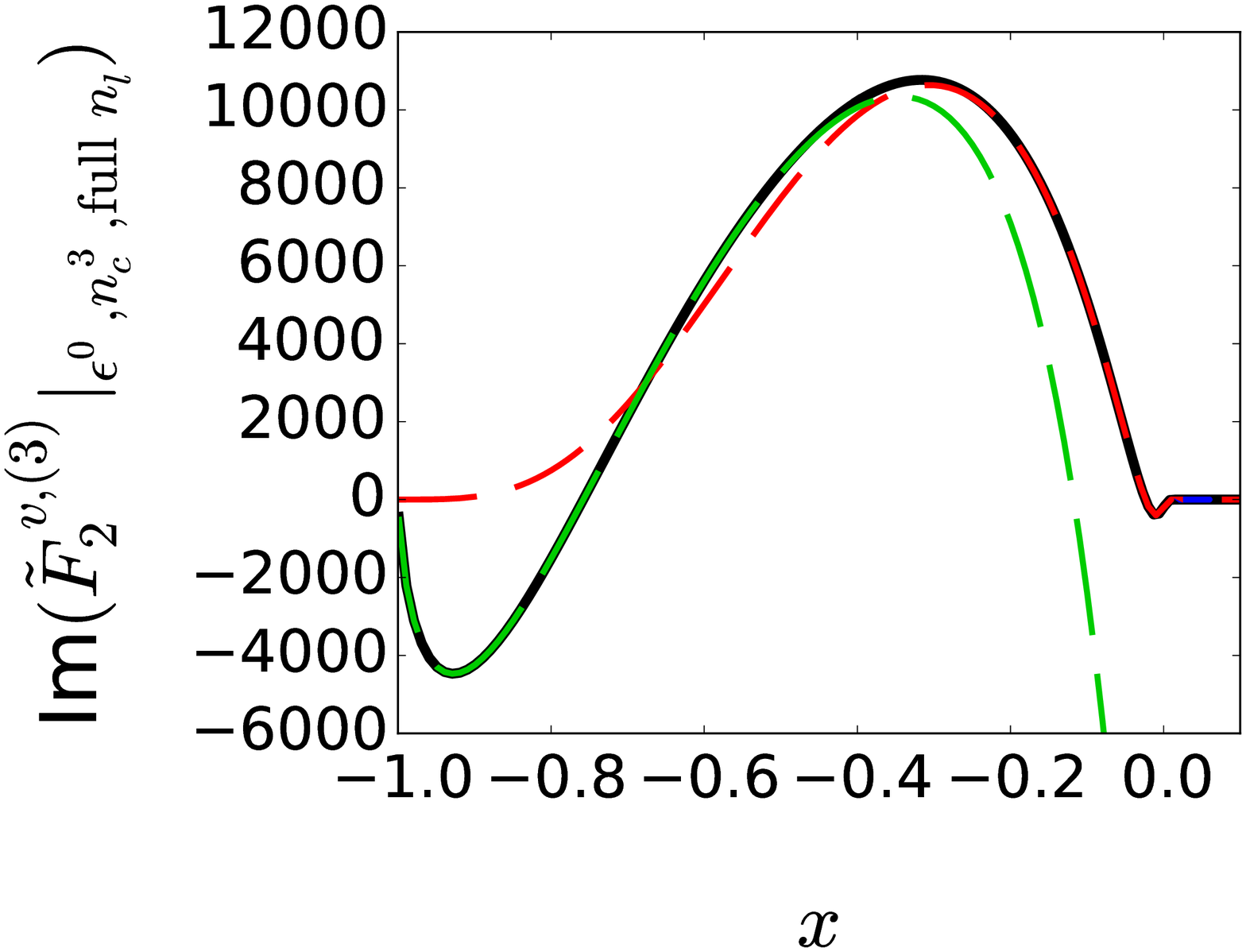} 
      \\
      \includegraphics[width=0.3\textwidth]{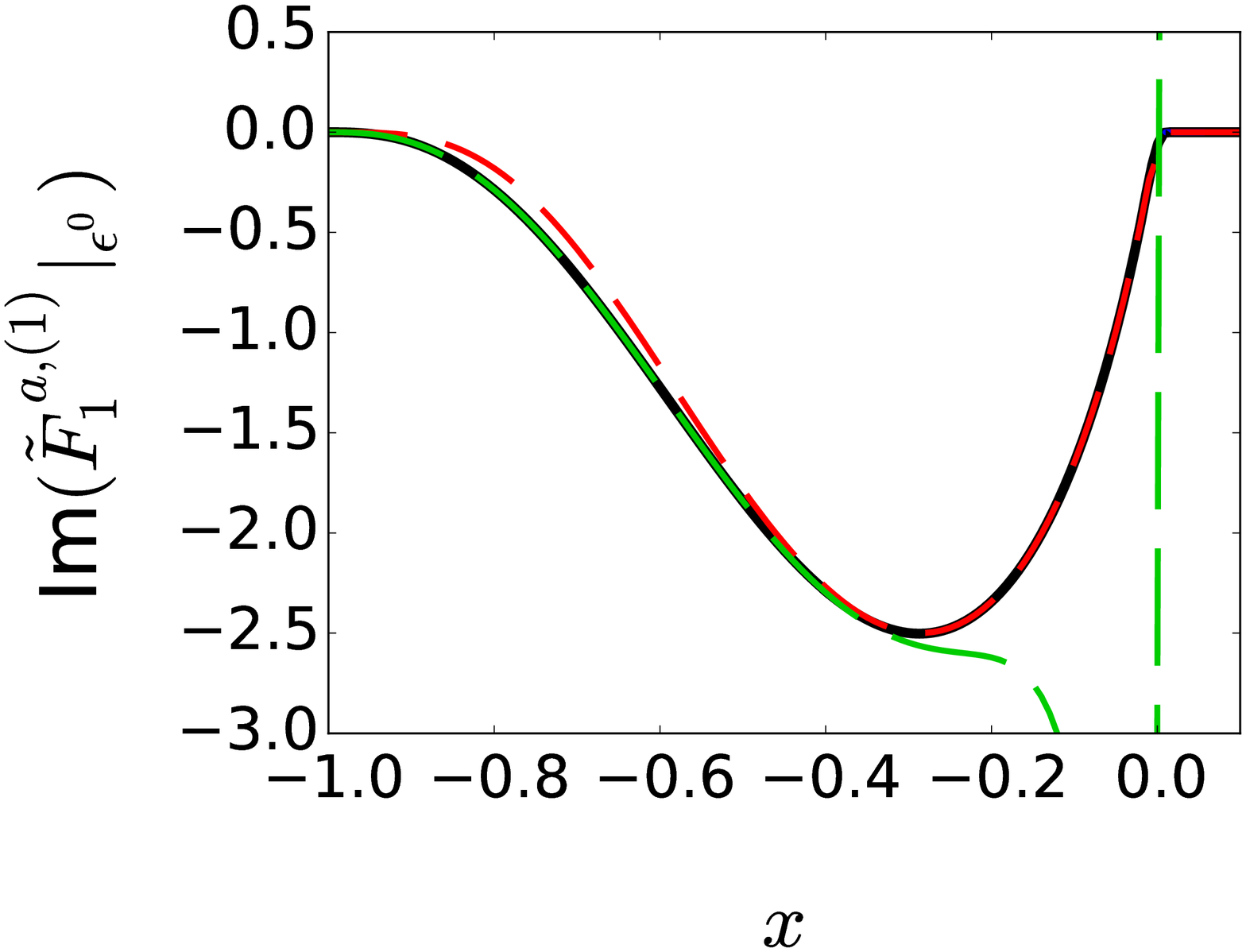} &
      \includegraphics[width=0.3\textwidth]{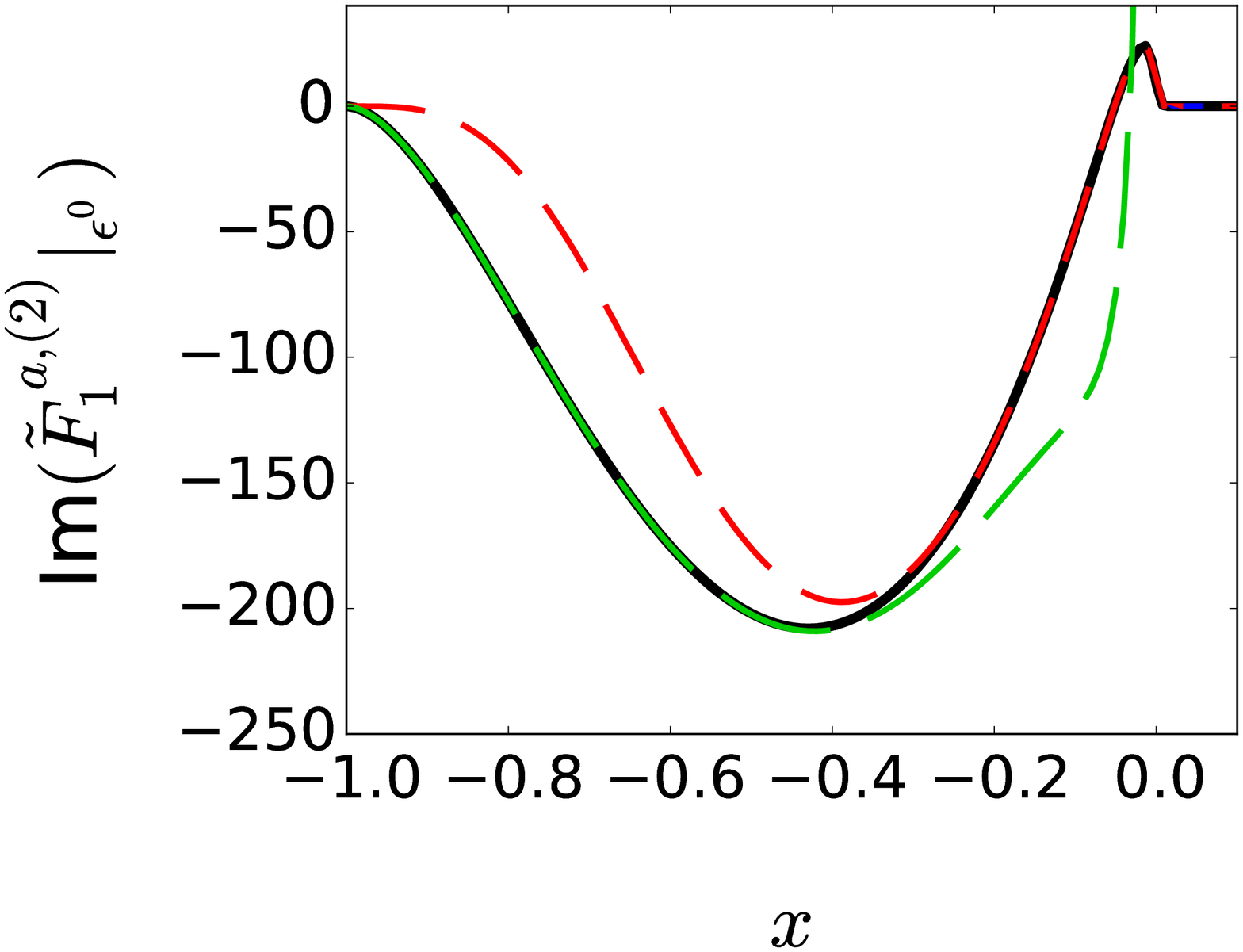} &
      \includegraphics[width=0.3\textwidth]{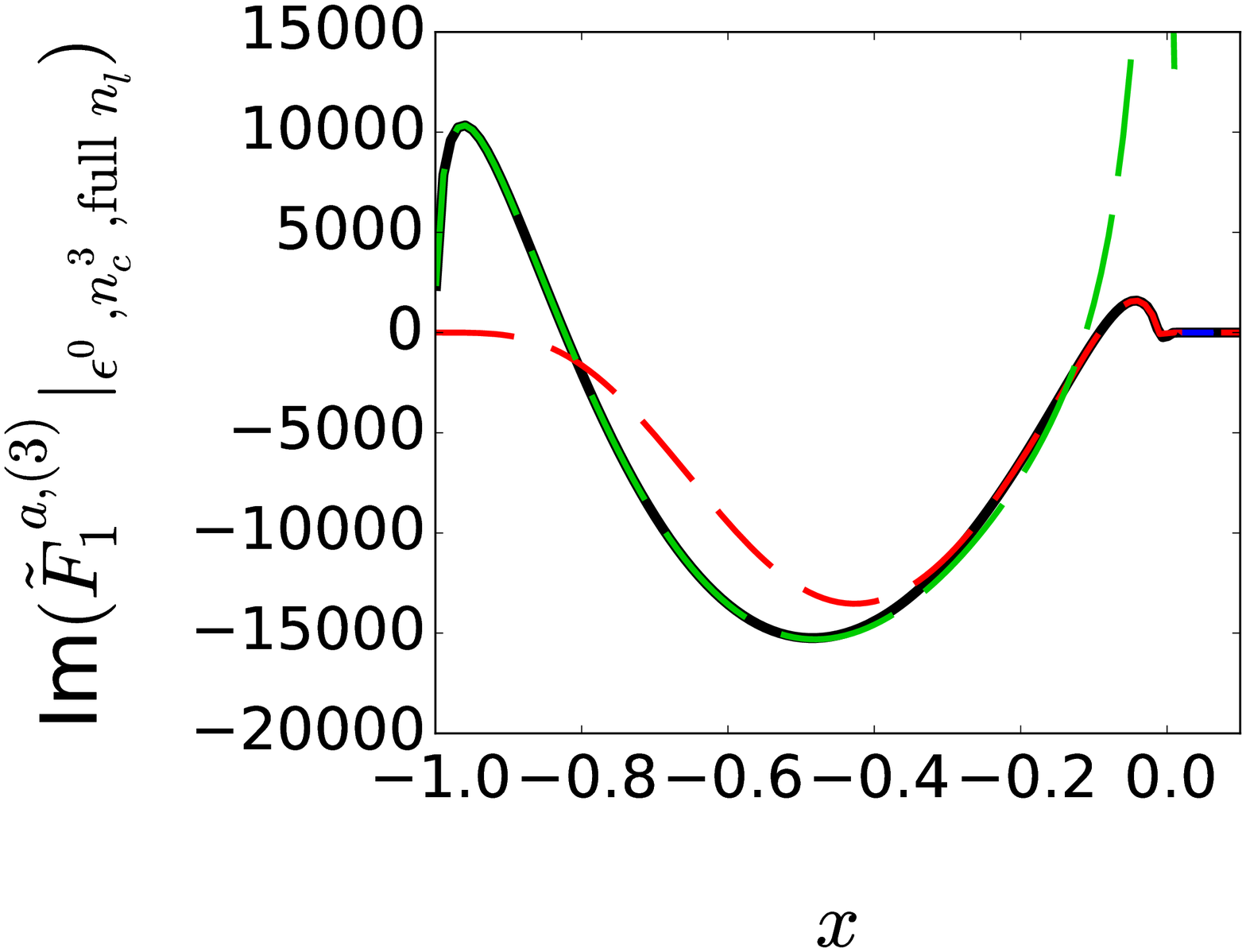} 
      \\
      \includegraphics[width=0.3\textwidth]{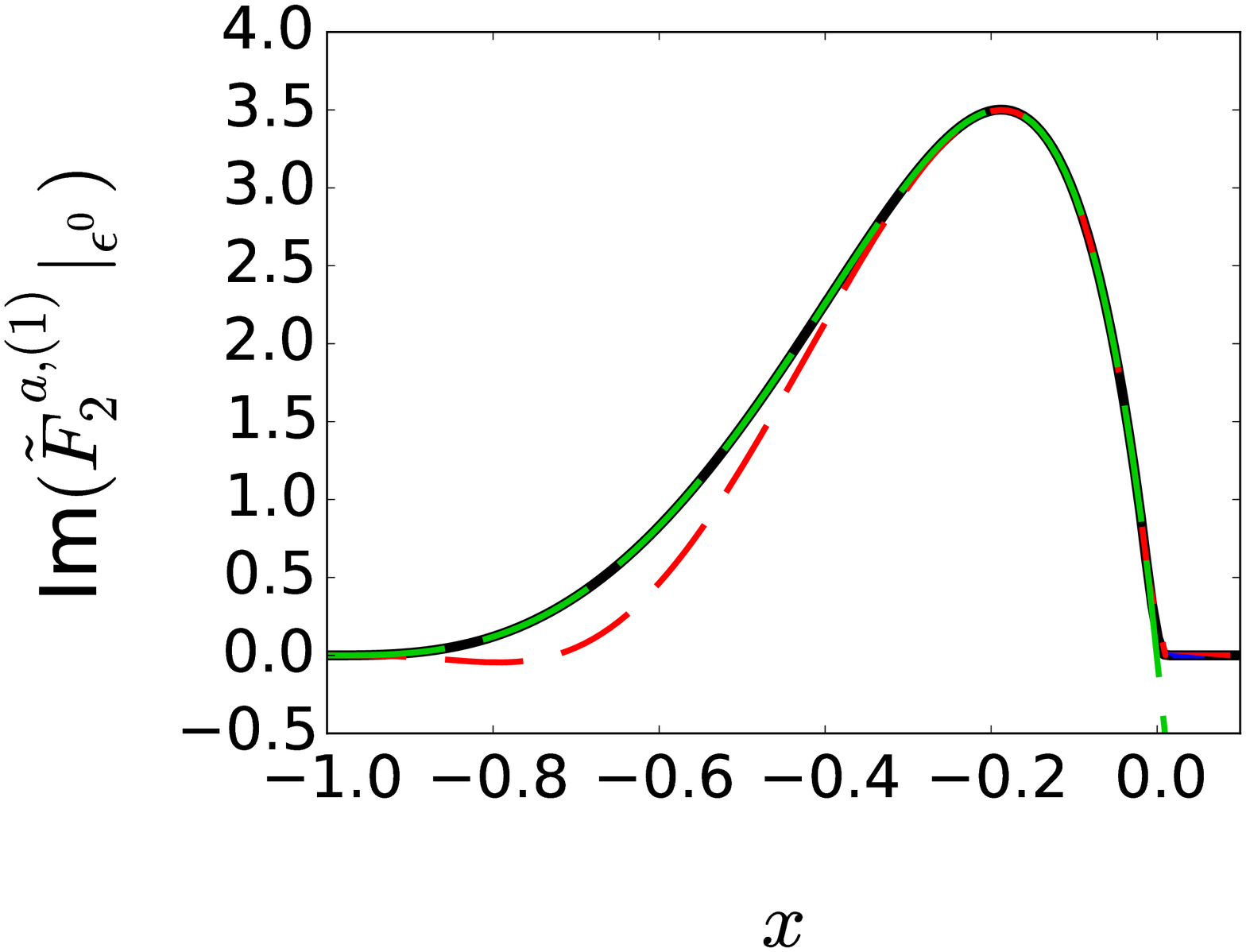} &
      \includegraphics[width=0.3\textwidth]{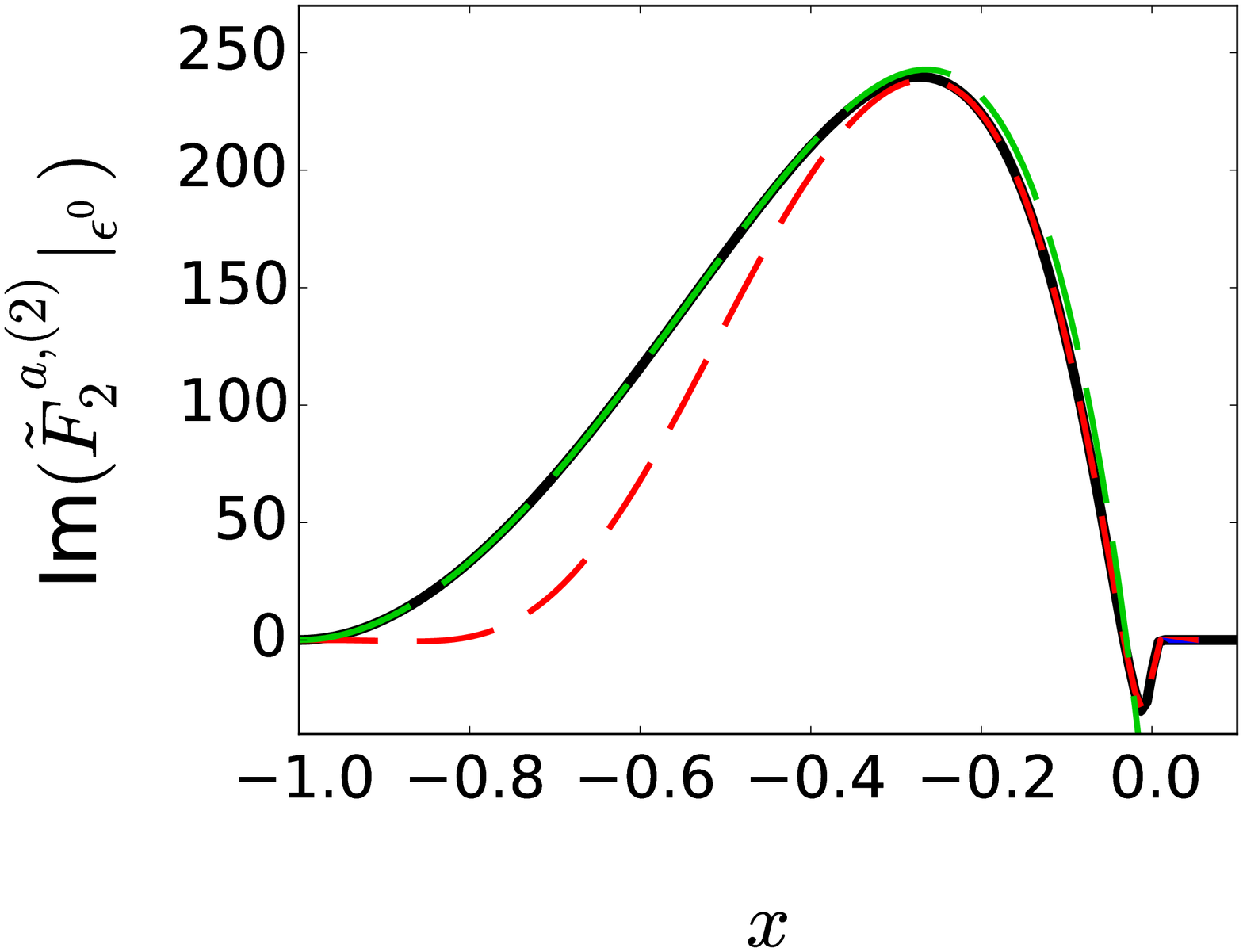} &
      \includegraphics[width=0.3\textwidth]{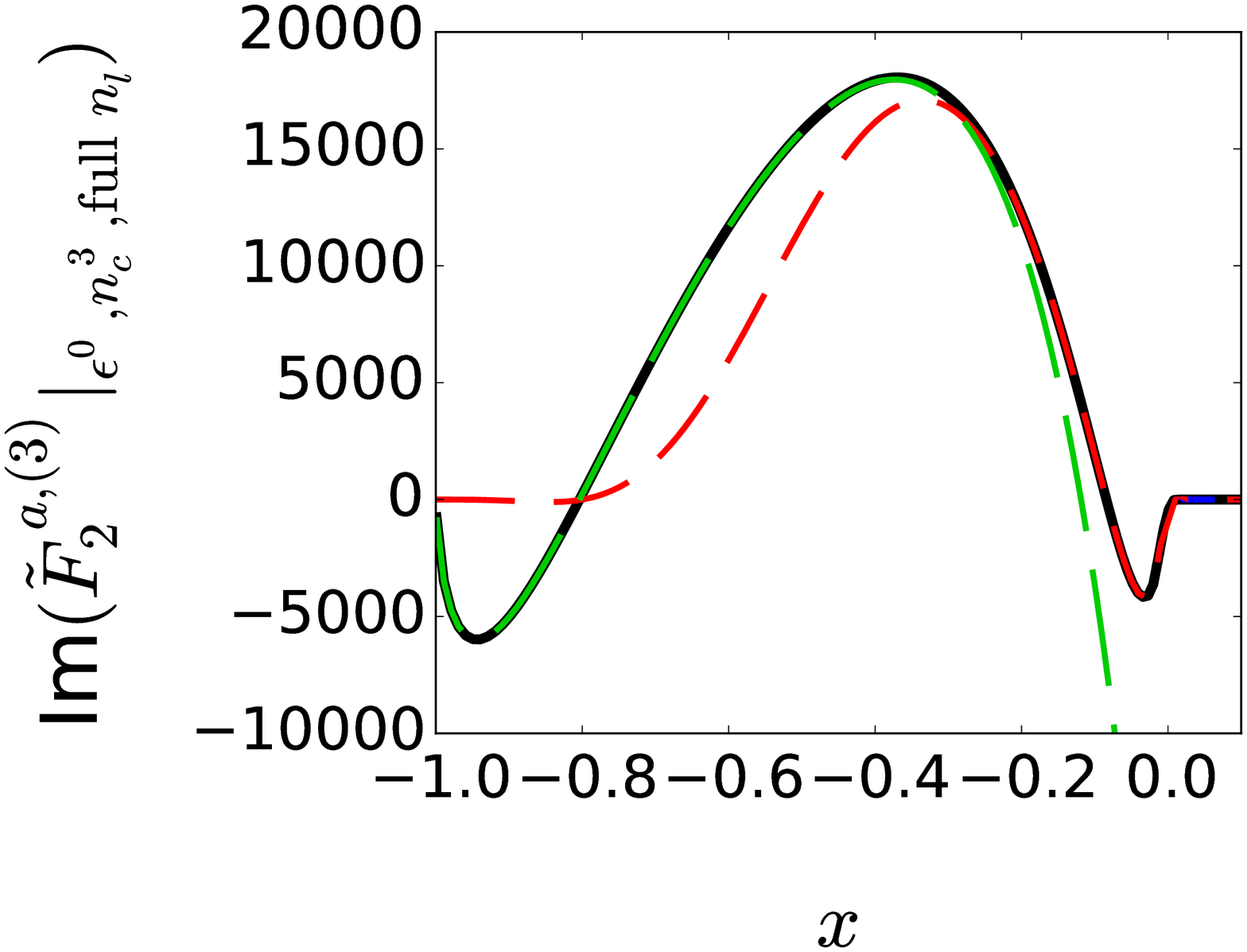} 
    \end{tabular}
    \caption{\label{fig::x_im_va}Imaginary part of the $\epsilon^0$ term of the
      vector and axial-vector form factors as a function of $x$. Exact results
      and approximations are shown as solid and dashed lines, respectively. At
      three-loop order we add the complete light-fermion part for $n_l=5$ and
      the $N_c^3$ contribution. Medium- (red) and long- (green)
      dashed lines correspond to the high-energy and threshold
      approximation, respectively. Note that the imaginary part is zero
      for $x\in[0,1]$.}
  \end{center}
\end{figure}

\begin{figure}[t] 
  \begin{center}
    \begin{tabular}{ccc}
      \includegraphics[width=0.3\textwidth]{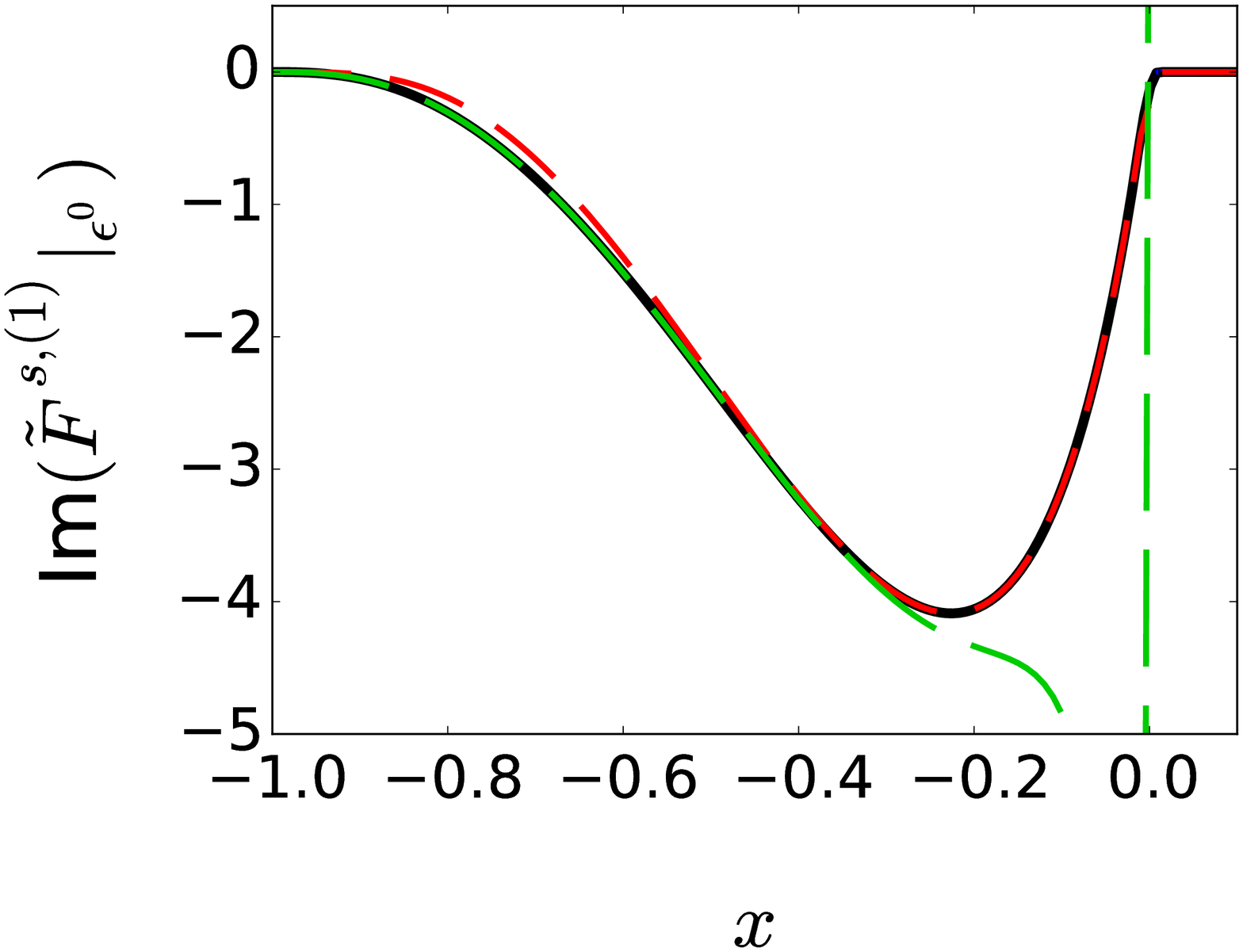} &
      \includegraphics[width=0.3\textwidth]{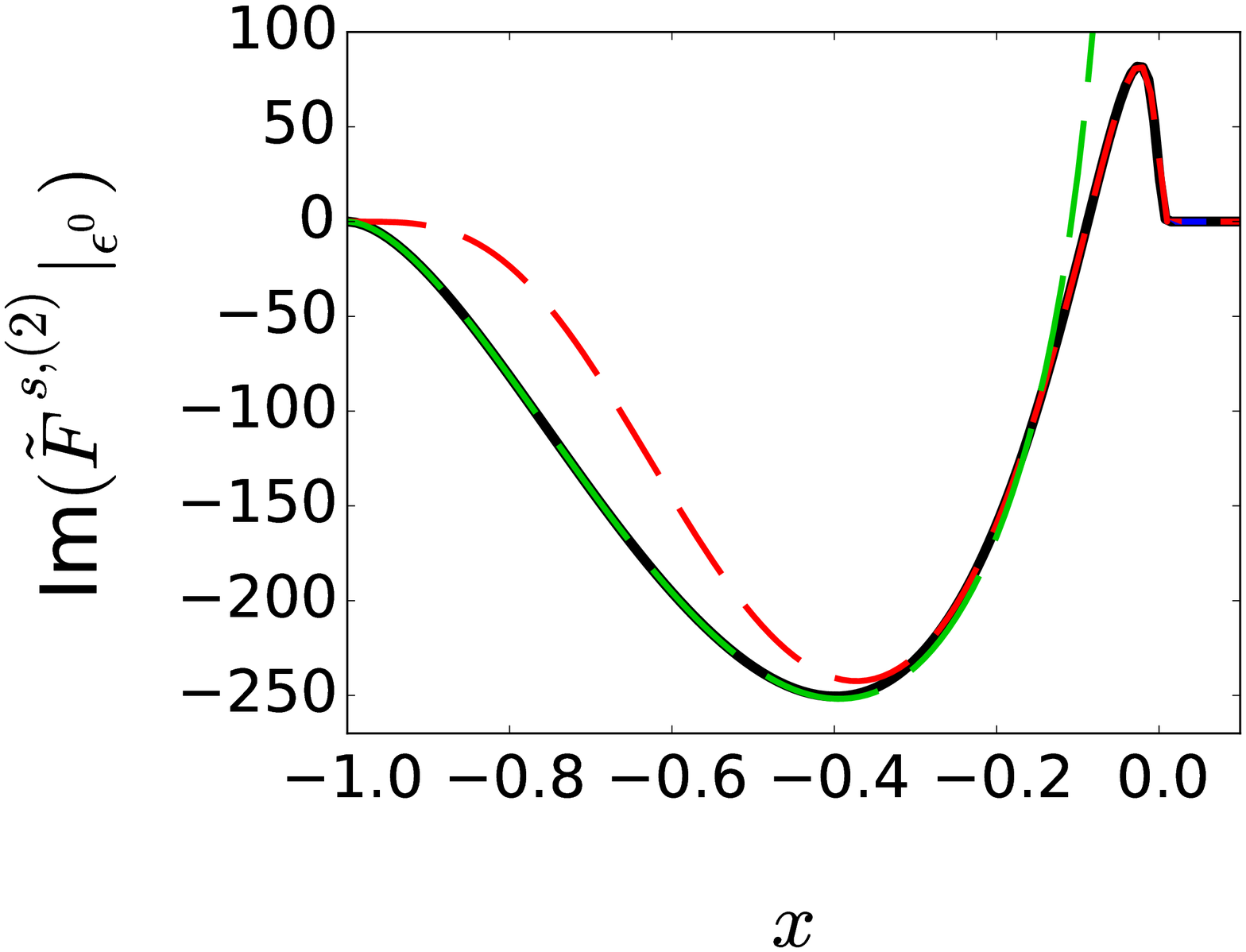} &
      \includegraphics[width=0.3\textwidth]{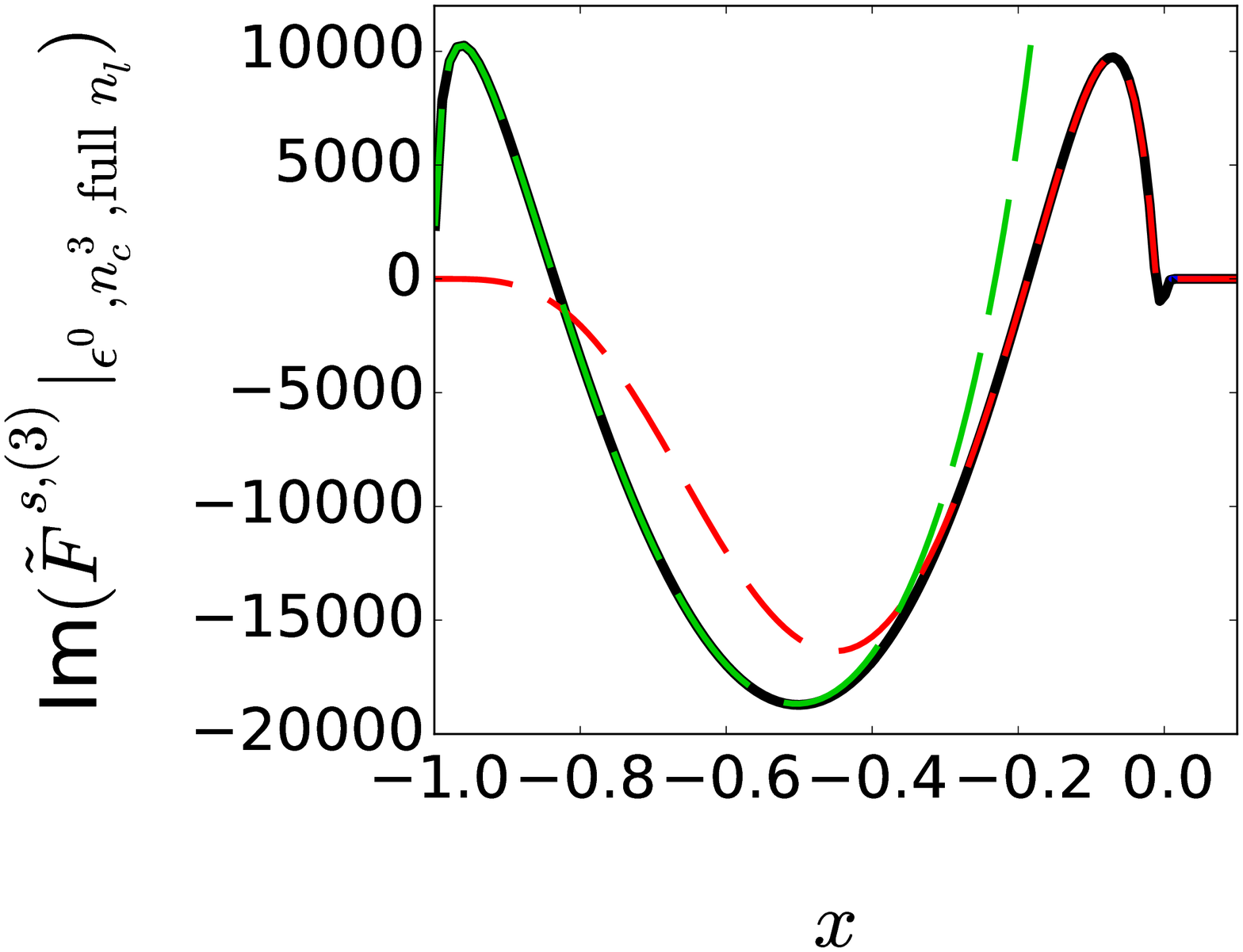} 
      \\
      \includegraphics[width=0.3\textwidth]{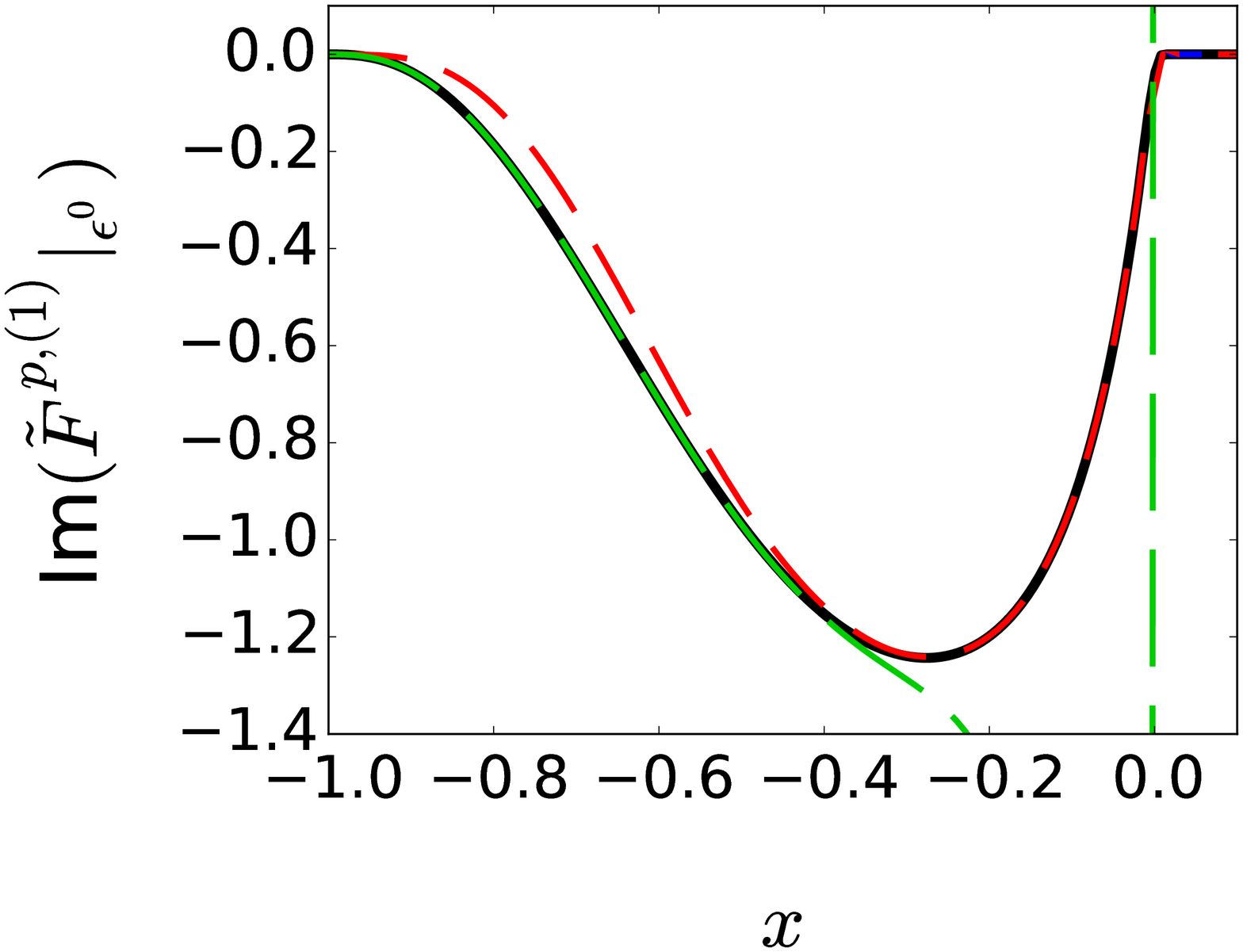} &
      \includegraphics[width=0.3\textwidth]{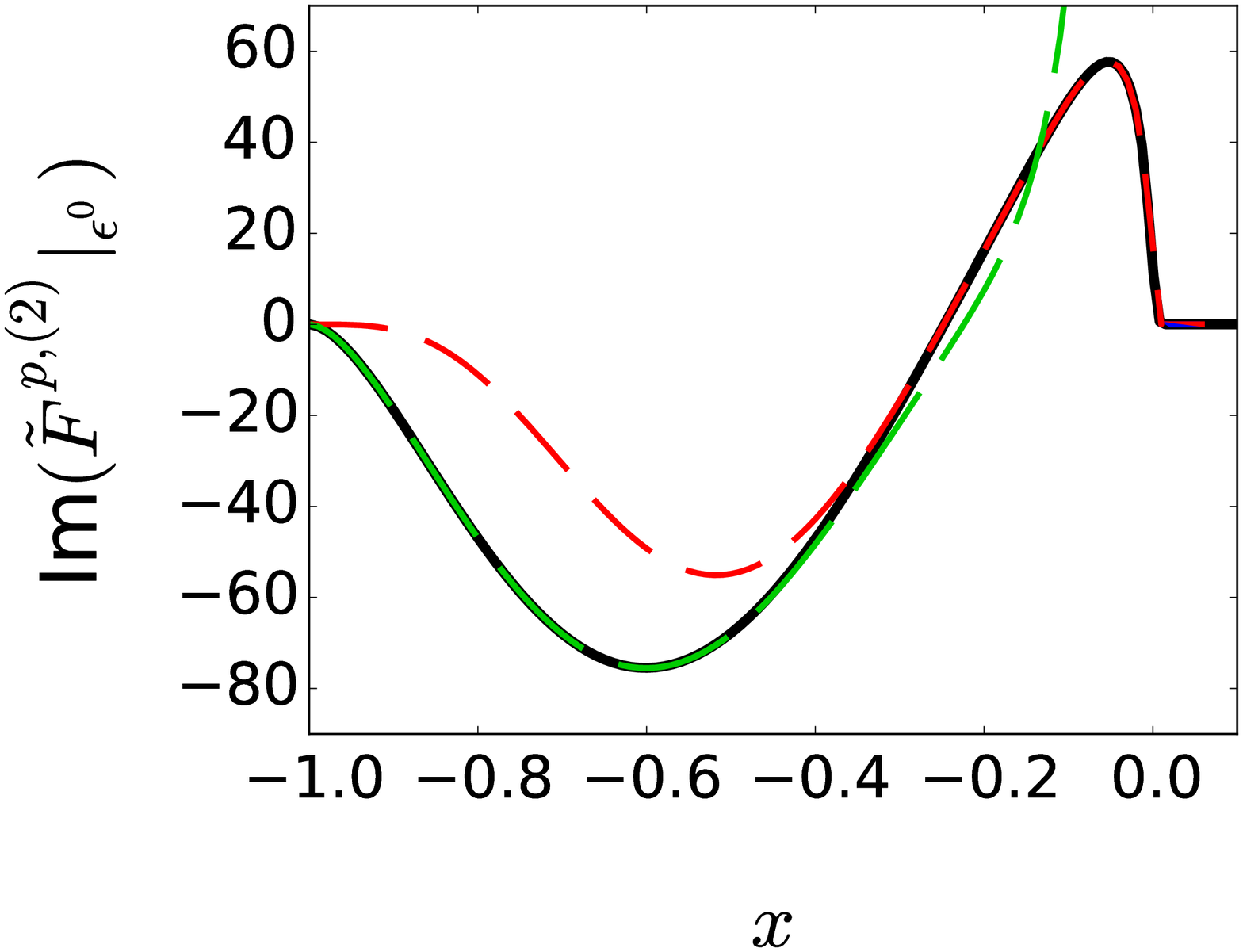} &
      \includegraphics[width=0.3\textwidth]{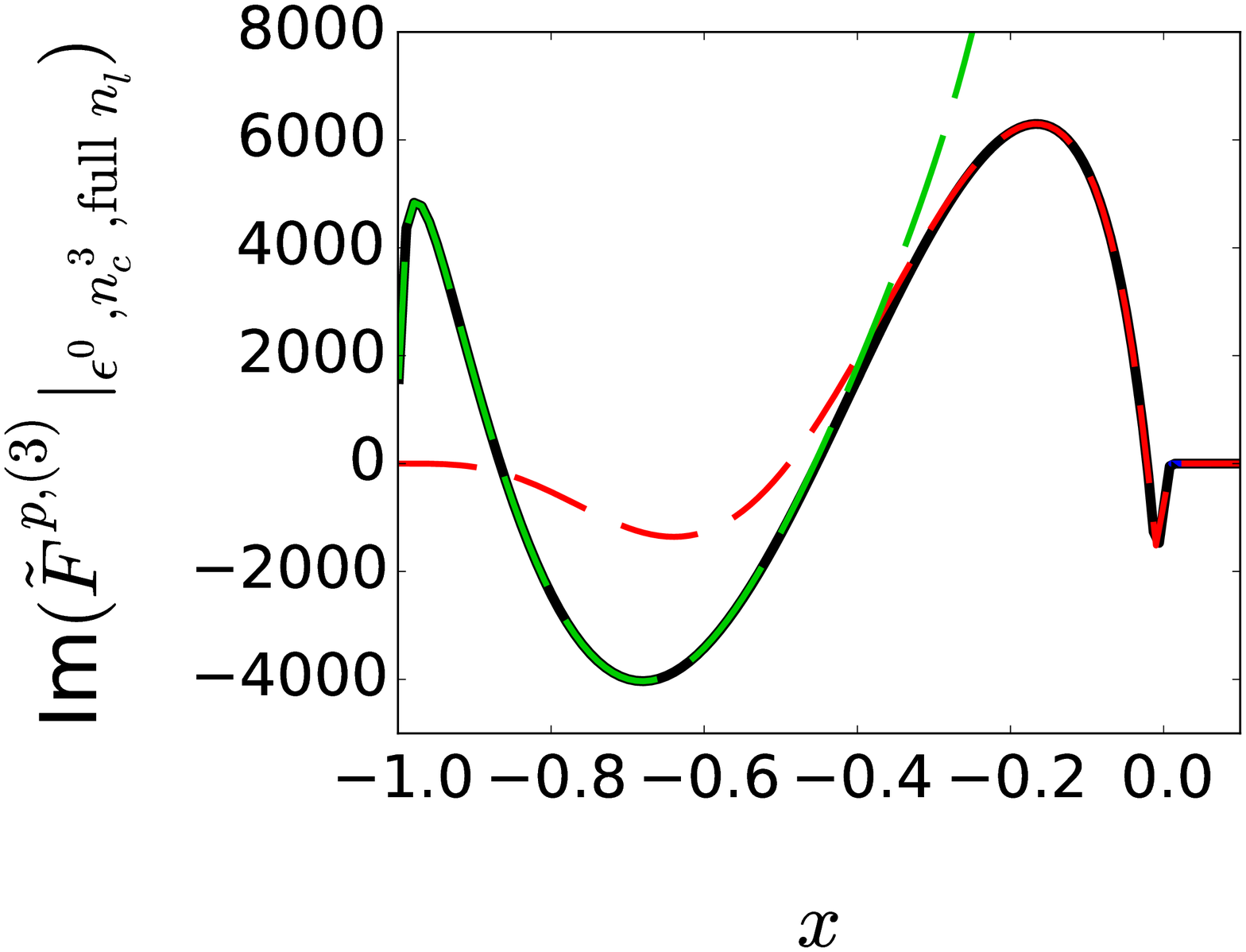} 
    \end{tabular}
    \caption{\label{fig::x_im_sp}Same as Fig.~\ref{fig::x_im_va} but for the
      the scalar and pseudo-scalar currents.}
  \end{center}
\end{figure}

We refrain to show analytic expressions but refer
to~\cite{Henn:2016kjz,Lee:2018rgs,Lee:2018rgs} for analytic results in the
kinematic limits $x\to0,1$ and $-1$.  In the following we complement the
numerical results shown in~\cite{Lee:2018rgs} by discussing the imaginary parts of
$F_1^v, \ldots, F^p$.

In~\cite{Lee:2018rgs} plots for the real part of the $\epsilon^0$ term of
\begin{eqnarray}
  \tilde{F}(q^2) = (1+x)^4 \left[ F(q^2) - F(q^2)\Big|_{q^2\to\infty} \right]
  \,.
  \label{eq::Ftil}
\end{eqnarray}
are shown for $x\in[-1,1]$. The subtraction term and the factor $(1+x)^4$
guarantee that in all three limis ($x\to0,1,-1$) finite results are obtained.
Furthermore, 
the six scalar functions $F_1^v, \ldots, F^p$ are multiplied by $(\pi-\phi)^4$
and plotted against $\phi\in[0,\pi]$, where $x=e^{i\phi}$. 

Note that for $x>0$ and for values of $x$ on the circumference of the unit
circle the form factors are real-valued. Thus, in Figs.~\ref{fig::x_im_va}
and~\ref{fig::x_im_sp} we show the imaginary parts of $\tilde F_1^v, \ldots,
\tilde F^p$ for
negative values of $x$. By construction (cf. Eq.~(\ref{eq::Ftil})) the form
factors are zero for $x=-1$ and $x=0$.  The exact result is shown as solid
black curve. Short- and long-dashed curves correspond to high-energy and
threhold approximations where terms up to order $x^6$ and $\beta^3$ [with $x =
2\beta/(1+\beta) - 1$] are included.  Note that in all cases (except for a
small region around $x\approx-0.3$ for the two-loop result of $F_1^v$) the
whole range $x\in[-1,1]$ can be covered by the approximations, i.e., for each
$x$-value there is at least one of the dashed curves on top of the (black)
solid line.


\section{Outlook}

There are several possible next steps towards the full massive three-loop form
factors. A well-defined and gauge invariant subset is the singlet
contributions where the coupling of the external current to the external
fermions is mediated via gluons. Another subset is composed of all
(non-singlet) contributions with a closed massive fermion loop.  However, all
these cases are significantly more involved which is mainly due to the
occurrence of so-called elliptic sectors where differential equations cannot
be transformed into a canonical form.  This makes it probably necessary to
resign on numerical methods, e.g., along the lines presented in
Ref.~\cite{Lee:2017qql}.


\section*{\label{sec::ack}Acknowledgments}

V.S. is thankful to Claude Duhr for permanent help in manipulations with GPLs.
This work is supported by RFBR, grant 17-02-00175A, and by the Deutsche
Forschungsgemeinschaft through the project ``Infrared and threshold effects in
QCD''.  R.L.  acknowledges support from the ``Basis'' foundation for
theoretical physics and mathematics.  The Feynman diagrams were drawn with the
help of {\tt Axodraw}~\cite{Vermaseren:1994je} and {\tt
  JaxoDraw}~\cite{Binosi:2003yf}.

\end{document}